\documentclass[12pt]{article}
\usepackage{geometry}
\usepackage{url}
\usepackage{latexsym}
\usepackage[centertags]{amsmath}
\usepackage{amssymb}
\usepackage{amsthm}
\usepackage{enumerate}
\usepackage{graphicx}
\usepackage{amsfonts,amssymb}
\usepackage{amsmath}
\usepackage[all]{xy}   
\usepackage{tikz-cd}

\usepackage{tikz}

\usepackage{tikz-cd} 
\usetikzlibrary{shapes.misc,arrows,decorations.markings}
\usetikzlibrary{matrix}
 
\newtheorem{Theorem}{Theorem}
\newtheorem{Definition}{Definition}
\newtheorem{Proposition}{Proposition}

\newcommand {\cC}{\mbox{${\mathcal C}$}}

\newcommand{\Der}{\mbox{${\mathrm{Der}}$}}

\newcommand{\calA}{\mbox{${\cal A}$}}

\newcommand{\calN}{\cal N}

\newcommand{\R}{\mathbb{R}}

\newcommand{\Cinf}{C^\infty }
\newcommand{\Mbar}{\mbox{$\bar{M}(\Lambda_n)$}}

\newcommand{\Mstar}{\mbox{$M_*$}}

\begin{document}

\title{Functorial Einstein Algebras and the Malicious Singularity}

\author{Michael Heller, Tomasz Miller \\ \normalsize Copernicus Center for Interdisciplinary Studies, Jagiellonian University \\ 
\normalsize Szczepa\'{n}ska 1/5, 31-011 Cracow, Poland \\ 
Leszek Pysiak, Wies{\l }aw Sasin\\ \normalsize Institute of Mathematics and Cryptology, Military University of Technology\\
\normalsize Kaliskiego 2, 00-908 Warsaw, Poland \\[12pt]}

\date{\today}
\maketitle

\begin{abstract}
Einstein algebra, the concept due to Geroch, is essentially general relativity in an algebraic disguise. We introduce the concept of Einstein--Grassmann algebra as a superalgebra (defining a supermanifold) which is also an Einstein algebra. We employ this concept to confront the supermanifold structure with the structure of strong singularity, the so-called malicious singularity, in general relativity. Einstein--Grassmann algebras consist of two parts: a part called body and a part called soul. For the body part, the singularity theorems apply and the singularities persist as the conclusions of the classical theorems on the existence of singularities require. We prove that, if we relax algebraical requirements, the soul part of the algebra can survive the malicious singularity. In particular, we study the behaviour of supercurves in the presence of malicious singularity. 
\end{abstract}

\section{Introduction}
Regardless of the fact that the recent LHC experiments have placed rather severe constraints on the existence of supersymmetry (SUSY) \cite{Canepa}, we should not forget that supergeometry already has a firm place in modern theoretical physics. From the spin--statistics theorem it follows that the intrinsic spin of a particle is strictly related to the statistics to which the particle is subject. Owing to this, the phase space of a field theory must be a superspace with even-graded coordinates corresponding to configurations of boson fields and odd-graded coordinates corresponding to those of fermion fields. The problem is not purely abstract, because it is the stability of matter that depends on quantum statistics \cite{Lieb}.

Einstein repeatedly expressed dissatisfaction that his field equations contained an `ugly' duality: on the left-hand side of the equations, the purely geometric gravitational part, on the right-hand side of them, the purely phenomenological description of matter by means of the energy--momentum tensor. It was Berezin who noted that this duality can be mitigated in supergeometric theories. 'If supersymmetric theories are taken more seriously, he wrote, it seems to me that they imply the existence of a fundamental symmetry between coordinates and fields' \cite[p. 28]{Berezin}. In these theories, the role of coordinates is largely taken over by even parts of Grassmann algebras, and generators of odd parts can be treated as classical equivalents of fermionic operators. This effect is even more pronounced if we require that the superalgebra defining the supermanifold under consideration is also an Einstein algebra, essentially general relativity in an algebraic dressing. In that case we will be talking about an Einstein--Grassmann superalgebra. Its definition and examination of its basic properties is one of the goals of this paper. The idea of Einstein  algebra was first proposed by Geroch \cite{Geroch72}, and then developed by two of the present authors \cite{Heller92,HelSasin94}.

The second goal of the present work is a confrontation of two things: the supermanifold structure and the structure of strong singularity (the so-called malicious singularity). The first of them serves to the construction of the theory of supergravity, the second is related to the theorems about the existence of classical singularities and remains till the present day one of unsolved problems in cosmology and relativistic astrophysics. The natural scale for supergravity is the Planck scale. Since this scale is currently experimentally unavailable, we develop a hypothesis that the conditions for the existence of singularities can possibly be treated as a kind of ``proxy information'' about what happens when the laws of physics, as we know them, break down after crossing the Planck threshold. 

The confrontation of the singularity problem with the supermanifold structure turned out to be fruitful. The Grassmann algebras, that are an essential part of this structure, consist of two parts: a part called body and a part called soul. The body is responsible for the macroscopic structure of space-time along with Einstein's equations; the soul is responsible for what happens at the microscale or even at the Planck level. Consequently, for the body part of the Grassmann algebra the classical singularity theorems apply and the singularities exist as their conclusions require, but the soul part of the Grassmann algebra is less sensitive to the existence of singularities and some parts of it pass through them, under certain conditions which are formulated below (Proposition \ref{supersingularity}). This applies even to the strongest singularities, called malicious singularities. We also study the behaviour of supercurves when they hit the singularity.

This is the objective of our work, but we are getting there gradually, using the method initiated in papers \cite{ROMP1,ROMP2}. The method basically consists in that that instead of considering a space, for example space-time $(M, \Cinf(M))$, where $\Cinf (M)$ is an algebra of smooth functions on $M$, we consider the space $(\bar{M}(A), \Cinf(\bar{M}(A)))$, where $A$ is a certain (not necessarily commutative) unitary algebra, $\bar{M}(A)$ a set of all algebra homomorphisms from $\bar{M}(A)$ to $A$, and $\Cinf(\bar{M}(A))$ the algebra of smooth functions on $\bar{M}(A)$. Although $\bar{M}(A)$ has ``more points'' than $M$, the structures on them, i.e. $\Cinf(M)$ and $\Cinf(\bar{M}(A))$, are isomorphic. In the language of category theory, $A$ plays the role of a stage in a suitable category. This provides an additional degree of generality to our considerations: substituting different algebras for $A$, we get different versions of the theory under construction. This also justifies the name functorial differential spaces. We show that if we construct a sheaf of algebras $\bar{\mathcal{A}}$ over $M(A)$ in such a way that $\bar{\mathcal{A}}(U_i) \cong \Cinf(\bar{U}_i) \otimes \Lambda $, where $U_i$ is an open subset in $M(A)$ and $\Lambda $ is a Grassmann algebra, we obtain a functorial version of the differential superspace, in the special case of a supermanifold. 

In the case of the Grassmann algebra ($A = \Lambda $), additional difficulties arise due to the fact that this algebra has an even (commutative) and an odd (noncommutative) part, so functions for functional analysis cannot be taken from $\Cinf $. The standard way, in supermanifold theory, is to define a class of functions, called $G^\infty $, appropriate to supermanifolds (see \cite{Rogers}). This is a very restrictive method, because it selects only those functions that behave appropriately on both the commutative and non-commutative parts (so they must meet very restrictive conditions). Our method is more tolerant: it consists in extending the class of admissible functions to all linear functions so that the admissible functions include both commutative and non-commutative parts. Since the Einstein--Grassmann algebra is defined with respect to the part of the Grassmann algebra called its body (the Grassmann algebra is an Einstein--Grassmann algebra if its body is an Einstein algebra), the conclusions regarding it obtained by our method are the same as those obtained by the  method  of $G^\infty $ functions (both methods overlap on the body).

Our road map leads our research as follows. In Section 2, we briefly recall two structures pivotal for our work: the Einstein algebra and the Grassmann algebra, and in Section 3 we get them to interact with each other by creating the concepts of Einstein--Grassmann functorial algebras and Einstein--Grassmann functorial differential spaces, correspondingly.

We now pose the question: how does the Einstein--Grassmann functorial differential space (or the Einstein--Grassmann functorial algebra) behave under the strong curvature singularity? At the time when the existence of singularities was a hot topic in relativistic cosmology, various constructions of singular boundaries of space-time were proposed (for a review see \cite{Dodson}). Among them Schmidt's $b$-boundary \cite{Schmidt} seemed the most promising, and since problems, into which it was involved, have been overcome by the theory of differential spaces \cite{Structured}, we will use it in our analysis. The strongest singularity we encounter in relativistic cosmology and relativistic astrophysics, defined in terms of Schmidt's construction, is called malicious singularity. In Section 4, we collect those results on this singularity that will be essential for our further research.

The main part of this research is contained in Sections 5--7. In Section 5, we prove the proposition on the behaviour of Einstein--Grassmann differential spaces in the presence of a malicious singularity. If everything happens in the category of $\Cinf $-algebras, the supermanifold reacts to the occurrence of a malicious singularity in essentially the same way as an ordinary manifold. However, if we relax the requirements of smoothness and limit ourselves only to the category of linear spaces and their transformations, some parts of the ``soul'' can survive the singularity (Proposition \ref{supersingularity}). In Section 6, we analyse, in more detail, how various supercurves behave when hitting a malicious singularity. In Section 7, we extend these results to the case when an Einstein--Grassmann differential space is an Einstein--Grassmann supermanifold.

\section{Preliminaries}\label{section:Prel}
In this section, we briefly recall two structures fundamental to our work:  the Einstein algebra and the Grassmann algebra, which we will bring to active interaction in the following sections.

To define Einstein algebra, we first need to prepare some concepts (see \cite{Geroch,Heller92,HelSasin94}). Let us consider a pair $(V, R)$, where $R$ is a (not necessarily commutative) ring and $V$ a (left) $R$-module. We further assume that $V$ is a free $R$-module of rank $n+1$, i.e. that every basis of $V$ has $n+1$ elements, and that $g: V \times V \to R$ is a Lorentz metric on $V$. The triple $(V, R, g)$ is called a Lorentz module. The typical example of a Lorentz module is $(\textrm{Der}(C^\infty(M)),C^\infty(M), g)$, where $(M,g)$ is a Lorentz manifold and $\textrm{Der}(C^\infty(M))$ is the $C^\infty(M)$-module of derivations of the algebra $C^\infty(M)$.

The Lorentz module $(V, R, g)$ is called an Einstein algebra, if on $(V, R, g)$ the following conditions hold
\begin{align*}
(i) & \quad \mathrm{Ric} - \tfrac{1}{2}rg  + \Lambda g = 8 \pi T,
\\
\textrm{or} \quad (ii) & \quad \mathrm{Ric} = \Lambda g,
\end{align*}
where all symbols have their usual meaning (for details of this definition see \cite{ROMP2}). In simple terms, a Lorentz module $(V, R, g)$ is an Einstein algebra, if $g$ satisfies Einstein's equations. 

Our idea is to modify the  Einstein algebra structure so that it can work in a supermanifold environment. Since a supermanifold is a space modeled on Grassmann algebras, we must now refer to this notion.

An associative $\mathbb{R}$-algebra $\Lambda_n $ with unity is called a Grassmann algebra if there is a set of generators in it $1, \beta_1, \ldots , \beta_n$, $n \in \mathbb{N}$, such that
\[
1 \beta_i = \beta_i = \beta_i 1
\]
and
\[
\beta_i \beta_k + \beta_k \beta_i = 0.
\]
Every element $\alpha \in \Lambda_n$ can be written in the form
\[
\alpha = \sum_{k\geq 0} \sum_{i_1, \ldots , i_k} \alpha_{i_1 \ldots i_k} \beta_{i_1} \ldots  \beta_{i_k},
\]
Indices $i_1, \ldots , i_k$ must satisfy one of the following conditions: either $\alpha_{i_1 \ldots i_k}$ should be antisymmetric with respect to their indices, or the indices should form the increasing sequence $i_1 < i_2 < \ldots < i_k$. Obviously, $\beta_i$ are nilpotents ($\beta_i^2 = 0$). The above definition of Grassmann algebra can be rephrased in such a way as to allow an infinite number of generators. 

There is a unique algebra homomorphism $b : \Lambda_n \to \R$ which maps $1$ onto $1$ and all the other generators to zero. It is called the body map and its image the body of $\Lambda_n$. The linear map $s$ which maps $\Lambda_n$ onto its nilpotent elements is called the soul map and its image the soul of $\Lambda_n$.

The Grassmann algebra admits a gradation $\Lambda_n =\, ^0 \! \Lambda_n \oplus \, ^1 \!\Lambda_n$ where $^0\! \Lambda_n$ and $^1\! \Lambda_n$ are spanned by the products of even and odd numbers of generators $\beta_i$, respectively. The subspace ${}^0\Lambda_n$ is also a subalgebra of $\Lambda_n$. (For more on Grassmann algebras see \cite{Berezin,Rogers,DeWitt}.)

\section{Einstein functorial algebras} \label{sec:SuperFunctorial}
In this section, we first introduce the functorial differential space concept,  i.e. the generalization of a differential space defined for any stage $A$, where $A$ is a not necessarily commutative algebra. Also for the generality reasons, we start by placing our considerations in the category $Lin$ of linear spaces and only then, if necessary, we move to the category $C^\infty$ of smooth algebras. Then, with the help of this newly introduced concept, we define the functorial Einstein algebra. 

Let $(M, \Cinf (M))$ be an $m$-dimensional differential manifold, and $A$ an $n$-dimensional linear space containing $\mathbb{R}$ as a subspace. By $\bar{M}_L(A) := [\Cinf (M), A]_{Lin}$ we will denote the set of linear maps
\[
\rho : \Cinf (M) \to A.
\]

For every smooth function $f \in \Cinf (M)$ we define the function $\bar{f}: \bar{M}_L(A) \to A$ by
\[
\bar{f}(\rho ) = \rho (f)
\]
for any $\rho \in \bar{M}_L(A)$.

It is easy to see that the evaluation $ev_p: \Cinf (M) \to \R , \, ev_p(f) = f(p)$, for any $p \in M $, belongs to $\bar{M}_L(A)$. The set of all functions $\bar{f}$, for $f \in \Cinf (M)$, will be denoted by $\Cinf(\bar{M}_L(A))$.

The mapping $J: \Cinf (M) \to \Cinf(\bar{M}_L(A))$ given by $J(f) = \bar{f}$
is a bijection. Indeed, suppose $\bar{f} = \bar{g}$ for some $f, g \in \Cinf (M)$. This implies that $\bar{f}(ev_p) = \bar{g}(ev_p)$ for any $p \in M$. Therefore, $f(p) = g(p)$. In fact, $J$ is an isomorphism between $\Cinf (M)$ and $\Cinf(\bar{M}_L(A))$.

The isomorphism allows to endow $\Cinf (\bar{M}_L(A))$ with the structure of a $\Cinf $-algebra transported from $\Cinf (M)$
\[
\omega (\bar{f}_1, \ldots , \bar{f}_k) = \overline{\omega (f_1, \ldots , f_k)}
\]
where $f_1, \ldots , f_k \in \Cinf (M)$ and $\omega \in \Cinf (\R^k )$. In particular, taking $\omega(x,y) = x + y$ and $\omega(x,y) =xy$, we obtain the operations of addition and multiplication, respectively (cf. \cite{ROMP1})
\[
\bar{f}_1 + \bar{f}_2 = \overline{f_1 + f_2}, \qquad \bar{f}_1 \cdot \bar{f}_2 = \overline{f_1 f_2}.
\]
Thus defined addition is nothing but the pointwise addition in $A$. Indeed,
\[
(\bar{f}_1 + \bar{f}_2)(\rho ) = (\overline{f_1 + f_2})(\rho ) = \rho (f_1 + f_2) = \rho (f_1) + \rho (f_2) = \bar{f}_1(\rho ) + \bar{f}_2(\rho)
\]
for every $\rho \in \bar{M}_L(A)$. However, multiplication need not agree with the pointwise multiplication in $A$ (especially that the latter might even be undefined!). In general, one can only write
\[
(\bar{f}_1 \cdot \bar{f}_2)(\rho ) = (\overline{f_1 f_2})(\rho ) = \rho(f_1 f_2).
\]
Even if $A$ is an algebra, we cannot continue transforming the right-hand side of the above equality if $\rho $ is not multiplicative. However, if we now change from the category of linear spaces to the category of $\Cinf \, \R $-algebras and thus assume $\rho $ to be an algebra homomorphism, the operation $\cdot$ indeed becomes the pointwise multiplication
\[
(\bar{f}_1 \cdot \bar{f}_2)(\rho ) = \rho(f_1 f_2) = \rho (f_1) \rho (f_2) = \bar{f}_1(\rho ) \cdot \bar{f}_2 (\rho ).
\]

In this way, we obtain the ringed space $(\bar{M}_L(A), \Cinf(\bar{M}_L(A)))$, which can be considered either in the category of linear spaces (in which case we will decorate $M$ with the subscript $L$), or in the category of $\Cinf \, \R$-algebras (in which case we will drop the  subscript $L$). When $A$ is not specified, $(\bar{M}_L(A), \Cinf(\bar{M}_L(A)))$ will be called a functorial differential space. In the following, to make our considerations more specific, we will often choose Grassmann algebra as a stage; in this case we will talk about Grassmann differential space. The choice of such a stage is also dictated by the fact that after appropriate ``tensoring'' of the differential structure of the Grassmann differential space (see below Section \ref{sec::super}), it becomes a differential superspace (a generalization of the supermanifold).

Because of the isomorphism $J: \Cinf (M) \to \Cinf(\bar{M}_L(A))$ we can lift selected geometric objects from $(M, \Cinf (M))$ to the Grassmann differential space $(\bar{M}_L(\Lambda_n),\Cinf (\bar{M}_L(\Lambda_n)))$ (with the subscript $L$ or without it). For instance, for every $X \in \Der (\Cinf(M))$ we define $\bar{X} \in \Der (\Cinf(\bar{M}_L(\Lambda_n)))$ by
\[
\bar{X}\bar{f} = \overline{Xf}.
\]
Analogously, for a tensor $T: \Der^k(\Cinf(M)) \to \Cinf(M)$, we define 
$\bar{T}: \Der^k(\Cinf(\bar{M}_L(\Lambda_n))) \to \Cinf(\bar{M}_L((\Lambda_n)))$ by
\[ 
\bar{T}(\bar{X}_1, \ldots , \bar{X}_n) = \overline{T(X_1, \dots , X_n)} 
\]
for $X_1, \dots , X_n \in \Der(\Cinf (M))$. Other tensors can be lifted similarly.

It can be easily seen that any Einstein algebra in space-time $(M, g)$ can be lifted to the functorial space-time $(\bar{M}_L(A), \bar{g})$. We do this by first lifting the Lorentz module $(\Der(\Cinf (M)), \Cinf (M), g)$ on space-time to a functorial differential space, which gives $(\Der(\Cinf (\bar{M}_L(A))), \Cinf (\bar{M}_L(A)), \bar{g})$, then we notice that it satisfies Einstein's equations lifted to a functorial differential space, i.e.
\begin{align*}
(i) & \quad \overline{\mathrm{Ric}} - \tfrac{1}{2}\bar{r}\bar{g}  + \Lambda \bar{g} = 8 \pi \bar{T},
\\
\textrm{or} \quad (ii) & \quad \overline{\mathrm{Ric}} = \Lambda \bar{g},
\end{align*}

In this way we obtain the Einstein functorial algebra. (In what follows, we will drop the cumbersome subscript $L $.) If, in particular, we substitute the Grassmann algebra for $A$, we will talk about the Einstein--Grassmann algebra. Sometimes (if there is no risk of ambiguity) we will call the expression $\Cinf (\bar{M}(\Lambda_n))$ itself the Einstein--Grassmann algebra.

\section{Malicious singularity} \label{section:Malicious}
In this section, we will try to see how a Einstein--Grassmann differential space behaves when a strong curvature singularity is found in its structure. Since the key substructure in this case is a space-time manifold $M$ with a singular boundary $\partial{M}$, $M_* = M \cup \partial M $, let us start by recalling its properties.

At the time when the existence of singularities in cosmological models was a hot topic in relativistic cosmology, various constructions of singular boundaries of space-time were proposed, for instance: $g$-boundary \cite{Geroch}, $p$-boundary  \cite{Dodson}, essential boundary \cite{Clarke}, causal boundary \cite{KronheimerPenrose}. Among them Schmidt's $b$-boundary \cite{Schmidt} seemed the most promising, but it also ran into serious problems (see below). Since these problems seem to have been largely overcome by the theory of differential spaces \cite{Structured}, we will continue to use this construction.

Let us first recall the construction of a $b$-boundary $\partial_bM$ of space-time $M$. Let $O(M)$ be a connected component of the fibre bundle of orthonormal frames over $M$. A connection on $M$ induces a Riemannian metric $G$ on $O(M)$ in such a way as to make horizontal and vertical subspaces orthogonal. This metric is not unique, but all metrics obtained in this way are uniformly equivalent.  Therefore, the following steps of the construction do not depend on the choice of the embedding (for details see \cite{Schmidt,Dodson}).  Now, we define the differential structure (the sheaf of smooth functions) $\cC (O(M))$ on $O(M)$ as generated by projections $\pi_i: \R^n \to \R$ (restricted to $O(M)$) onto the $i$th coordinate. With the help of the distance function, defined in terms of $G$ , we construct the Cauchy completion $\overline{O(M)}$ of $O(M)$ (again as generated by projections). After extending the action of the structural group $O(3, 1)$ of the bundle from $O(M)$ to $\overline{O(M)}$, we form the quotient space $\overline{O(M)}/O(3, 1)$, and we define
$$
\partial_b M := \Mstar - M = \pi(\overline{O(M)}) - \pi (O(M))
$$
where $\pi: \overline{O(M)} \to \overline{O(M)}/O(3, 1)$ is the canonical projection. (For details of this construction see the original paper by Schmidt \cite{Schmidt} and \cite{Structured}.)

According to this construction, the elements of the $b$-boundary of space-time are not only the equivalence classes of causal geodesics, but also the equivalence classes of timelike curves of bounded acceleration\footnote{Timelike curves of unbounded acceleration are not supposed to represent any physical objects.} (i.e. geodesics and curves that could be interpreted as ``terminating'' at the singularity). For this reason, the $b$-boudary definition was considered the best definition of strong curvature singularity. This belief was put to an end by the discovery of Bosshard \cite{Bosshard}  and Johnson \cite{Johnson} that in the closed Friedman solution and the Schwarzschild solution of Einstein field equations if $p\in \partial_bM$ then the fiber $\pi^{-1}(p)$ of the fibre bundle of linear frames over $p$ degenerates to a single point which, in the case of the closed Friedman solution, is an obvious paradox (the beginning and the end of the universe are the same point). The situation has been clarified by the following theorem: 
\begin{Theorem}\label{b-Theorem}
If the fibre $\pi^{-1}(p), \; p \in \partial_bM$, in the fibre bundle of linear spaces $\overline{O(M)}$ over $M \cup \partial_bM$ degenerates to a single point then the only global cross-sections of the ringed space $(\Mstar, \Cinf_{M_*})$, where $\Cinf_{M_* }$ is the sheaf of smooth functions over $\Mstar $, are constant functions. Moreover, the only open (in the quotient topology) neighbourhood of $p$ is the entire $\Mstar $.
\end{Theorem}
\noindent \textbf{The proof} of the first part of this theorem uses the fact that if the fiber $\pi^{-1}(p)$ degenerates to a single point, then the boundary $\partial_bM$ is a single orbit of the action of the structural group $O(3, 1)$ on $\overline{O(M)}$, and in such a case any global section of $\Cinf(\overline{O(M)})$ is constant on orbits. The rest of the argument refers to the convergence of the sequences of reference frames to the boundary. The proof of the second part of the theorem is based on the fact that if $V$ is a neighborhood of $p \in \partial_bM$ then $\pi^{-1}(V)$ is an $O(3,1)$-invariant open set in $\overline{O(M)}$, i.e. if $V$ contains a point, it contains its orbit. And if $s: M \to O(M)$ is the cross-section of the sheaf, such that $s(M) \subset \pi^{-1}(V)$ then it must coincide with $\overline{O(M)}$. (For details of the proof see \cite{HelSasin94,Structured}). 

The singularities to which this theorem applies are also called malicious singularities. It is important to notice that $\Cinf_{M_*}|M = \Cinf (M)$, that is to say the sheaf $\Cinf_{M_*}|M$ contains enough functions to define the manifold structure of $M$, but of all these functions only the constant functions can be prolonged to the boundary $\partial_bM$. This situation can also be described in the following way. Topology on $\Mstar $ is given by
\[
\mathrm{top}\Mstar = \mathrm{top}M \cup \{\Mstar\}
\]
with the property that each two points $p, q \in M$ can be Hausdorff separated from each other, but no point $p \in M$ can be Hausdorff separated from $* \in \partial_bM$ since the only open set containing $*$ is the entire $\Mstar $. 

\section{Malicious singularity in an Einstein--Grassmann differential space} 
Let us consider space-time $M_* = M \cup \{*\}$ with malicious singularity $*$. The corresponding Einstein--Grassmann differential space is of the form $({\bar{M}_*} (\Lambda_n ), \Cinf ({\bar{M}_*}(\Lambda_n)))$.

On the strength of the previous section, however, $C^\infty(M_*)$ contains only the constant functions and so every $\rho \in {\bar{M}_*} (\Lambda_n ) = [C^\infty(M_*), \Lambda_n]_L$ is determined by its action on the function identically equal to one, $\rho(r\mathbf{1}) = r \rho(\mathbf{1})$ for any $r \in \R$.

Let us decompose $\rho$ into its ``body'' and ``soul'' with the help of the body and soul maps, namely
\begin{align*}
\rho = \rho_0 + \rho_S, \quad \textnormal{where} \quad \rho_0 = b \circ \rho \quad \textnormal{and} \quad \rho_s = s \circ \rho.
\end{align*}
It turns out that if we work in the category $\Cinf$, i.e., if we additionally assume that $\rho$ is multiplicative, then $\rho_s$ must vanish.

Indeed, by the multiplicativity of $\rho$ we have $\rho(\mathbf{1})^2 = \rho(\mathbf{1})$ and hence also $\rho_0(\mathbf{1})^2 = \rho_0(\mathbf{1})$, because the body map $b: \Lambda_n \rightarrow \R$ is an algebra homomorphism. We have thus two possibilities: $\rho_0(\mathbf{1}) = 0$ or $\rho_0(\mathbf{1}) = 1$.

If $\rho_0(\mathbf{1}) = 0$, then we obtain that $\rho_s(\mathbf{1})^2 = \rho_s(\mathbf{1})$. But since the soul of an element is nilpotent (cf. \cite[Proposition 3.1.2]{Rogers}), namely $\rho_s(\mathbf{1})^{n+1} = 0$, we must have
\begin{align*}
\rho_s(\mathbf{1}) = \rho_s(\mathbf{1})^2 = \rho_s(\mathbf{1})^3 = \ldots = \rho_s(\mathbf{1})^{n+1} = 0.
\end{align*}
Notice that in this case $\rho$ is the zero map.

As for the second case, $\rho_0(\mathbf{1}) = 1$ implies that $(1 + \rho_s(\mathbf{1}))^2 = 1 + \rho_s(\mathbf{1})$, what reduces to $\rho_s(\mathbf{1})^2 = -\rho_s(\mathbf{1})$. But again invoking the nilpotence $\rho_s(\mathbf{1})^{n+1} = 0$, also this time we obtain
\begin{align*}
\rho_s(\mathbf{1}) = -\rho_s(\mathbf{1})^2 = \rho_s(\mathbf{1})^3 = \ldots = (-1)^n \rho_s(\mathbf{1})^{n+1} = 0.
\end{align*}
In this case, $\rho$ is thus simply $\rho(r\mathbf{1}) = r$ for any $r \in \R$.

Therefore, if we work in the category $\Cinf$, malicious singularity annihilates the soul of the Grassmann algebra. In this case we do not get anything new compared to the situation in space-time with a malicious singularity (Theorem \ref{b-Theorem}).

Let us now consider the case when everything happens in the category of linear spaces; consequently $\rho $ need not be multiplicative. Since only constant functions prolong to the malicious singularity, we have
\[
\rho(r\mathbf{1}) = r\rho_0(\mathbf{1}) + r \sum_{k \geq 1} \sum_{{i_1}, \ldots , i_{k}} \rho_{{i_1} \ldots i_{k}}(\mathbf{1})\beta_{i_1} \ldots \beta_{i_{k}},
\]
for any $r \in \R$. But now $\rho_{i_1 \dots i_k}: \Cinf (M) \to \R $ are only linear operators, and they can all be non-vanishing. In fact, there is a natural isomorphism ${\bar{M}_*} (\Lambda_n ) = [C^\infty(M_*), \Lambda_n]_L \cong \Lambda_n$, which assings $\Lambda_n \ni \lambda \mapsto \rho_\lambda$ with the laSincetter map defined via
\[
\rho_\lambda(r\mathbf{1}) = r\lambda, \quad r \in \R.
\] 
Therefore, as we can see, in the category of linear spaces much more elements survive the singularity thanks to employing $\Lambda_n$ as the stage.

We can summarize the above results in the form of the following proposition:

\begin{Proposition} \label{supersingularity}
Let $({\bar{M}_*}(\Lambda_n), \Cinf(\bar{M}_*(\Lambda_n)))$ be an Einstein--Grassmann differential space of dimension $(m, n)$ containing malicious singularity, ${\Mstar} = M \cup \{*\}$. In the case when $\rho $ is multiplicative, $\rho_0: \Cinf (M_*) \to \R, \rho_0(r\mathbf{1}) = r$, and $\rho_s: \Cinf (M_*) \to \Lambda_S, \, \rho_s(r\mathbf{1}) = 0$. In the case when $\rho $ is not required to be multiplicative, it is of the form $\rho(r\mathbf{1}) = r\lambda$ for any chosen $\lambda \in \Lambda_n$. In particular, $\rho_s$ can then be different from zero in the singularity. $\Box $
\end{Proposition}

It should not be forgotten that the appearance of a malicious singularity in a manifold $M$ is associated with the degeneracy of topology (see the end of Section \ref{section:Malicious}).

To further investigate what happens near the singularity, we should consider the behavior of various types of curves (supercurves) when they try to pass through the singularity. This will be done in the next sections. Let us notice that the conditions of the above proposition guarantee that the classical theorems on the existence of singularities remain valid (when we remain in the category $\Cinf $ and restrict to the corresponding body).

\section{Curves} \label{sec:supercurves} 
In this section, we study curves in an Einstein--Grassmann differential space and their behaviour when they hit malicious singularity. In defining such curves we follow \cite[pp. 67--68]{Rogers} with slight modifications required by our formalism. The analyzes of this section are valid for both $\rho $ multiplicative and nonmultiplicative.

Let $(M, \Cinf (M))$ be an $m$-dimensional smooth manifold, regarded as (a subset of) the body of an Einstein--Grassmann differential space $(\bar{M}(\Lambda_n ), \Cinf (\bar{M}(\Lambda_n)))$, and let $\gamma : I \to M$ be the usual curve in $M$. Let us show how it can be naturally lifted to the curve (on the stage $\Lambda_n$)
\[
\bar{\gamma }: \bar{I}(\Lambda_n) \to \Mbar.
\]

To this end, for any $\tau \in \bar{I}(\Lambda_n)$ define 
\[
\bar{\gamma} (\tau ) = \tau \circ \gamma^*,
\]
where $\gamma^*: C^{\infty }(M)^\ast \rightarrow C^{\infty }(I)^\ast$ is the pullback of $\gamma$. The construction can be visualised in the form of the following commutative diagram
\begin{center}
\begin{tikzcd}
C^{\infty }(I) \arrow[rr, "\tau"]                                           & {} \arrow[d, "\bar{\gamma}", bend right, shift right=4] & \Lambda_n \\
                                                                            & {}                                                  &           \\
C^{\infty }(M) \arrow[uu, "\gamma^*"] \arrow[rruu, "\tau \circ \gamma^* "'] &                                                     &          
\end{tikzcd}
\end{center}
Observe that $\bar{\gamma} (\tau ) \in \bar{M}(\Lambda_n)$ acts on any given $f \in C^\infty(M)$ via
\[
\bar{\gamma} (\tau )(f) = (\tau \circ \gamma^*)(f) = \tau(f \circ \gamma).
\]
In the special case of $\tau = ev_t$ for some $t \in I$, we obtain that
\[
\bar{\gamma} (ev_t)(f) = ev_t(f \circ \gamma) = f(\gamma(t)) = ev_{\gamma(t)}(f)
\]
or simply $\bar{\gamma}(ev_t) = ev_{\gamma(t)}$. It is in this sense that $\bar{\gamma}$ is a lifting of $\gamma$.

Let us follow the usage, common is supermanifold theory to take into account the gradation $\Lambda_n =\, ^0 \! \Lambda_n \oplus \, ^1 \!\Lambda_n$ (see Section \ref{section:Prel}), and to consider curves relating to even and odd variables separately. Accordingly, we have three types of supercurves:
\begin{enumerate}
\item
curves of the type $\bar{\gamma}_{1,0}$, when we take into account only even part of $\Lambda_n $
\[
\bar{\gamma}_{1,0}: \bar{I}(^0\! \Lambda_n) \to \bar{M}(^0 \! \Lambda_n),
\]
\item 
curves of the type $\bar{\gamma }_{0,1}$, when we take into account only odd part of $\Lambda_n$
\[
\bar{\gamma}_{0,1}: \bar{I}(^1\! \Lambda_n) \to \bar{M}(^1 \! \Lambda_n).
\]
Notice that curves of this type make sense only if we work in the $Lin$ category (because $^1 \! \Lambda_n$ is not an algebra!)
\item
curves of the type $\bar{\gamma }_{1,1}$, when we take into account the whole of $\Lambda_{n}$
\[
\bar{\gamma}_{1,1}: \bar{I}(\Lambda_n) \to \bar{M}(\Lambda_n).
\]
\end{enumerate}

The behaviour of these curves in the malicious singularity is determined by Proposition \ref{supersingularity}. We will return to this issue in the next section.

\section{Einstein--Grassmann supermanifolds}
\label{sec::super}
There were functorial differential spaces that constituted the environment of our considerations in previous sections. We examined a number of their properties and formulated some conclusions regarding the occurrence of malicious singularities in them. They served as a kind of research tool rather than as models of physical spaces. In this section, we will show that a slight modification of their differential structure (tensor multiplication by a Grassmann algebra) will transform them into (somewhat generalized) supermanifolds (or, more generally, super differential spaces).

There are at least several (not all equivalent) definitions of a supermanifold: for instance, Berezin--Leites original definition \cite{Berezin,BerezinLeites}, Kostant's definition \cite{Kostant}, DeWitt's definition in terms of maps and atlases \cite{DeWitt}. In what follows, we will refer to Berezin's definition (as it was formulated by Rogers \cite[p. 86]{Rogers}) for its simplicity. The definition runs as follows.
\begin{Definition} \label{def:supermanifold}
A smooth real (algebro-geometric) supermanifold of dimension $(m, n)$ is a pair $(M, \calA )$ where $M$ is a real $m$-dimensional manifold and $\calA $ a sheaf of supercommutative algebras over $M$ satisfying the following conditions
\begin{enumerate}
\item
there exists an open cover $\{U_i|U_i \in \mathrm{top}M, i \in I \}$ of $M$ such that for each $i \in I$
\[
{\mathcal A} (U_i) \cong C^{\infty } (U_i) \otimes \Lambda_n,
\]
where $\Lambda_n$ is a Grassmann algebra  with $n$ generators,
\item
if $\calN $ is the sheaf of nilpotents in $\calA $, then
\[
(M, {\mathcal A} /{\mathcal N}) \cong (M, C^{\infty }_M)
\]
where $\Cinf_M$ is a sheaf of smooth functions on $M$.
\end{enumerate}
\end{Definition}
We will call the supermanifold defined in this way the Berezin supermanifold. Let us notice that an $(m, 0)$-dimensional supermanifold is simply an $m$-dimensional manifold.

The adaptation of the above definition to the environment of functorial supermanifolds is straightforward. When we work in the $\Cinf $ category, all we need to do is add the bars  above the appropriate symbols. When we work in the category of linear spaces, the only decision that has to be made is the choice of the appropriate topology needed to define the counterpart of the sheaf ${ \mathcal A}$. We propose the following strategy.

Let $(M, \Cinf (M))$ be a differential manifold, and $(\bar{M}_L(A), \Cinf (\bar{M}_L(A))$ its functorial counterpart. The functor ${\cal F} : U \to \Cinf (U)$, where $U \in \mathrm{top} M$, is a sheaf on $M$. We form another sheaf on $M$ by definig the functor $\bar{\cal{F}}: U \to\Cinf(\bar{U}_L (A))$, where $\bar{U}_L = [\Cinf(U), A]_{Lin}$. It can be easily seen that sheaves $\cal{F}$ and $\bar{\cal{F}}$ are isomorphic (let us notice that for the functions $\bar{f}_i$ over $(U_i)_{i \in I}$ we have $\bar{f}|_{U_i} = \overline{f|_{U_i}}$). The above, with $A = \Lambda_n$, leads to the following definition.
\begin{Definition} \label{def:functsupermanifold}
A smooth real functorial (algebro-geometric, Berezin) supermanifold of dimension $(m, n)$ is a pair $(\bar{M}_L(\Lambda_n), \bar{\calA})$ where $\bar{M}_L(\Lambda_n) = [\Cinf(M), \Lambda_n]_{Lin}$, with $M$ real $m$-dimensional manifold, $\Lambda_n $ the Grassmann algebra with $n$ generators, and $\bar{\calA}$ a sheaf of supercommutative algebras over $M$ satisfying the following conditions
\begin{enumerate}
\item
there exists an open cover $\{U_i|U_i \in \mathrm{top}M, i \in I \}$ of $M$ such that for each $i \in I$
\[
\bar{{\mathcal A}} (U_i) \cong C^{\infty } ((\bar{U}_i)_L) \otimes \Lambda_n,
\]
\item
if $\bar{\calN}$ is the sheaf of nilpotents in $\bar{\calA}$, then
\[
(M, \bar{{\mathcal A}} /\bar{{\mathcal N}}) \cong (M, C^{\infty }_M)
\]
where $\Cinf_M$ is a sheaf of smooth functions on $M$.
\end{enumerate}
\end{Definition}

Taking into account the isomorphism of differential structures $\Cinf (M)$ and $\Cinf(\bar{M}_L(\Lambda_n))$, we see that Definition \ref{def:functsupermanifold} is only a slight modification of Definition \ref{def:supermanifold}; it is, in fact, a different formal expression of the same mathematical structure. However, owing to this modification, we can see  that our previous results concerning the malicious singularity also apply to the Berezin supermanifold.

Let us consider two functorial differential spaces, $(\bar{M}(A), \Cinf (\bar{M}(A)))$ and $(\bar{N}(A), \Cinf (\bar{N}(A)))$, and let us notice that multiplication by a tensor does not spoil the smoothness. The mapping
\[
\bar{F}: (\bar{M}(A), \Cinf (\bar{M}(A))) \to (\bar{N}(A), \Cinf (\bar{N}(A))),
\]
given by $\bar{F}(\rho ) = \rho \circ F^*, \, \rho \in \bar{M}(A)$, is smooth, if for every $\beta \in \Cinf (N)$, one has $\bar{\beta} \circ \bar{F} \in \Cinf(\bar{M}(A))$. It can be easily checked that in our case
\[ 
\bar{F}: (\bar{M}(A), \Cinf (\bar{M}(A)) \otimes \Lambda_n) \to (\bar{N}(A), \Cinf (\bar{N}(A)) \otimes \Lambda_n),
\]
the analogous condition is satisfied. Indeed, for any element of $\Cinf (\bar{N}(A)) \otimes \Lambda_n$, being of the form $\sum_i \bar{\beta}_i \otimes \lambda_i$, one has that
\begin{align*}
\left( \sum\nolimits_i \bar{\beta}_i \otimes \lambda_i \right) \circ \bar{F} = \sum\nolimits_i \left( \bar{\beta}_i \circ \bar{F}\right) \otimes \lambda_i,
\end{align*}
what is a well-defined element of $\Cinf (\bar{M}(A)) \otimes \Lambda_n$.

The concept of an Einstein--Grassmann algebra from the end of Section \ref{sec:SuperFunctorial} is automatically transferred to the current situation. We will call the sheaf $\bar{\mathcal{A}}$ the Einstein--Grassmann sheaf if, for every $\bar{U}_i$, $\bar{\mathcal{A}}(\bar{U}_i) \cong \Cinf(\bar{U}_i) \otimes \Lambda_n$, the algebra $\Cinf(\bar{U}_i)$ is an Einstein algebra.
A supermanifold with an Einstein--Grassmann algebra as its differential structure will be called Einstein--Grassmann supermanifold.

We should now investigate how the Einstein--Grassmann supermanifold behaves when space-time $M$ has a malicious singularity $*$, $M_* = M \cup \{*\}$. Instead of considering the sheaf $\bar{\mathcal{A}}$, it is enough to consider the set of its global sections $\Cinf (\bar{M}_\ast(\Lambda_n)) \otimes \Lambda_n$. In what follows, one can work in the $\Cinf $ category or in the category of linear spaces, as appropriate. Since $C^\infty(M_\ast) = \R\mathbf{1}$, we have that
\[
(\bar{M}_\ast(\Lambda_n), \Cinf (\bar{M}_\ast(\Lambda_n)) \otimes \Lambda_n) \cong (\bar{M}_\ast(\Lambda_n), \R \mathbf{1} \otimes \Lambda_n) \cong ( \bar{M}_\ast(\Lambda_n), \Lambda_n).
\]
As we can see, only constants of the Grassmann algebra $\Lambda_n$ survive the prolongation to the singular boundary. However, let us look at Proposition \ref{supersingularity}. Its proof does not change if the structural algebra is tensored by $\Lambda_n$. Therefore, in the case with multiplicativity ($\Cinf $ category), $\rho_s$ vanishes, and only a single $\rho_0$ remains in $\bar{M}_\ast(\Lambda_n)$, just like in the case without tensoring by $\Lambda_n$. For the case with no multiplicativity (category of linear spaces), this limitation does not hold, $\rho_s$ need not vanish, and $\bar{M}_\ast(\Lambda_n) \cong \Lambda_n$. Of course, all topological peculiarities remain in force when passing to the malicious singularity.

When studying the singularity problem, one should consider curves hitting the singularity. However, it is easy to see that tensor multiplication by $\Lambda_n$ does not change the results obtained in Section \ref{sec:supercurves}. Indeed, let us consider the supercurve
\[
\bar{\gamma}: (\bar{I}(\Lambda_n), \Cinf(\bar{I}(\Lambda_n)) \otimes \Lambda_n) \to (\bar{M}(\Lambda_n), \Cinf(\bar{M}(\Lambda_n)) \otimes \Lambda_n).
\]
Like any curve, a supercurve $\bar{\gamma}$ is a mapping from the parameter space $\bar{I}(\Lambda_n)$ to its target space $\bar{M}(\Lambda_n)$, while tensor multiplication by $\Lambda_n$ affects only the structure algebra $\Cinf(\bar{M}(\Lambda_n)) \otimes \Lambda_n$. We conclude that tensor multiplication by $\Lambda_n$ does not significantly change the results as far as the appearance of malicious singularity is concerned.

\medskip

As we have seen, the description of a malicious singularity in an Einstein--Grassmann differential space is not significantly different from its description in space-time. This is because the effects of a malicious singularity in an Einstein--Grassmann differential space are largely determined by the effects of this singularity in its body, i.e. in space-time. In fact, the only difference consists in allowing for the case of non-multiplicativity. This has permitted the appearance of quantities that do not disappear in Einstein--Grassmann differential spaces when a malicious singularity is present. 

A more important property than the disappearance or non-disappearance of certain quantities in the presence of singularities seems to be the pathology associated with the change in topology. This pathology affects both space-times and Einstein--Grassmann differential spaces equally. As long as we stay in the open subset (of space-time or of an Einstein--Grassmann differential space), everything functions as it should. However, as soon as we try to extend our analysis to the singularity, everything degenerates into one open subset and all the functions that we can define on it are constant functions. It is as if the singularity controls everything (its only open neighbourhood is entire $M_\ast$), although space-time locally, at a safe distance from the singularity (in an open neighbourhood), retains its independence.

\end{document}